\documentclass[a4paper,12pt]{article}

\usepackage[english]{babel}

\usepackage[a4paper,top=2cm,bottom=2cm,left=3cm,right=3cm,marginparwidth=1.75cm]{geometry}

\usepackage{amsmath}
\usepackage{amsfonts}
\usepackage{graphicx}
\usepackage{tikz}
\usetikzlibrary{quantikz2}
\usepackage[colorlinks=true, allcolors=blue]{hyperref}
\usepackage{glossaries}

\newacronym{TN}{TN}{Tensor Networks}
\newacronym{QC}{QC}{Quantum Computing}
\newacronym{STN}{STN}{Singularity Tensor Network}
\newacronym{GPR}{GPR}{Gaussian Process Regressor}
\newacronym{MAE}{MAE}{Mean Absolute Error}
\newacronym{TT}{TT}{Tensor Train}
\newacronym{TCA}{TCA}{TT-cross approximation}
\newacronym{LSMC}{LSMC}{Least Square Monte Carlo}

\title{STN-GPR: A Singularity Tensor Network Framework for Efficient Option Pricing}

\begin{document}
\maketitle

\begin{center}
    \begin{tabular}{c}
        Dominic Gribben\textsuperscript{1,*}, Carolina  Allende\textsuperscript{1}, Alba Villarino\textsuperscript{1,*},\\ Aser Cortines\textsuperscript{2,*}, Mazen Ali\textsuperscript{1}, Román Orús\textsuperscript{1, 3, 4}
    \end{tabular}
    \\
    \vspace{0.5em}
    \small{
    \begin{tabular}{c}
        \textit{\textsuperscript{1}Multiverse Computing, Paseo de Miram\'on 170, 20014 San Sebasti\'an, Spain} \\
        \textit{\textsuperscript{2}Multiverse Computing, rue de la Croix Martre, 91120 Palaiseau, France} \\
        \textit{\textsuperscript{3}DIPC, Paseo Manuel de Lardizabal 4, E-20018 San Sebastián, Spain}\\
        \textit{\textsuperscript{4}Ikerbasque Foundation for Science, Maria Diaz de Haro 3, E-48013 Bilbao, Spain}\\
        \textsuperscript{*}\textit{\{dominic.gribben, alba.villarino, aser.cortines\}@multiversecomputing.com}
    \end{tabular}
    }
\end{center}

\vspace{1em}

\begin{center}
    \begin{tabular}{c}
        Pascal Oswald\textsuperscript{1,*},
        Noureddine Lehdili \textsuperscript{1,*}
    \end{tabular}
    \\
    \vspace{0.5em}
    \small{
    \begin{tabular}{c}
        \textit{\textsuperscript{1}Natixis CIB}\\
        \textsuperscript{*}\textit{\{pascal.oswald, noureddine.lehdili, \}@natixis.com}
    \end{tabular}
    }
\end{center}
  
\vspace{2.5em}

\begin{abstract}
We develop a tensor-network surrogate for option pricing, targeting large-scale portfolio revaluation problems arising in market risk management (e.g., VaR and Expected Shortfall computations). The method involves representing high-dimensional price surfaces in tensor-train (TT) form using TT-cross approximation, constructing the surrogate directly from black-box price evaluations without materializing the full training tensor. For inference, we use a Laplacian kernel and derive TT representations of the kernel matrix and its closed-form inverse in the noise-free setting, enabling TT-based Gaussian process regression without dense matrix factorization or iterative linear solves. We found that hyperparameter optimization consistently favors a large kernel length-scale and show that in this regime the GPR predictor reduces to multilinear interpolation for off-grid inputs; we also derive a low-rank TT representation for this limit. We evaluate the approach on five-asset basket options over an eight-dimensional parameter space (asset spot levels, strike, interest rate, and time to maturity). For European geometric basket puts, the tensor surrogate achieves lower test error at shorter training times than standard GPR by scaling to substantially larger effective training sets. For American arithmetic basket puts trained on LSMC data, the surrogate exhibits more favorable scaling with training-set size while providing millisecond-level evaluation per query, with overall runtime dominated by data generation.
\end{abstract}

\section{Introduction}
\label{sec:intro}
The price of an option, like that of any financial instrument, is ultimately determined by supply and demand. There is, however, the notion of a “fair” option price, obtained by considering various market conditions and imposing a set of modelling assumptions. The most famous and widely used framework of this kind is the Black–Scholes model, which leads to the Black–Scholes partial differential equation. The tractability of this approach depends strongly on the type of option under consideration: for European options written on a single asset, or on a basket with a geometric payoff, closed-form solutions exist and prices can be computed at negligible computational cost~\cite{black1973,merton1973,kemna1990,stulz1982}. For most other options, however, analytic solutions are unavailable and numerical methods are required~\cite{wilmott1995,brennan1977}. In many applications, particularly in risk management, a large number of valuations is needed to forecast potential losses, which can lead to substantial computational costs depending on the option type~\cite{glasserman2004mc,broadie2004mesh,lehdili2019market, lehdili2025performance}. The goal of this paper is to present an alternative to existing numerical approaches, focusing specifically on the pricing of American options.

In many practical applications, for example in risk management, the central computational task is to revalue large derivatives portfolios under thousands to millions of simulated market scenarios. These scenario-based portfolio revaluations are needed to compute tail risk measures such as Value at Risk (VaR) and Expected Shortfall (ES), as well as for stress testing and the determination of regulatory capital requirements. In this setting, the computational burden comes not from a single pricing call, but from repeatedly valuing an entire portfolio across a large number of scenarios with high dimensional risk factor shocks.

American-style features introduce additional challenges: early exercise leads to an optimal stopping problem, path dependence can arise, and the relevant state space often spans many risk factors (e.g. spot, volatility, rates) and contract parameters (e.g. strikes, maturities, barriers). Classical numerical techniques such as finite differences, lattice methods, and Monte Carlo with regression (e.g., LSMC), are effective but can become prohibitively expensive when many revaluations are needed for risk, calibration, or scenario analysis~\cite{brennan1977,cox1979,longstaff2001,jaillet1990,broadie2004mesh}. Surrogate models aim to amortize this cost by learning a mapping from inputs to prices through a limited number of calls to a trusted pricer~\cite{resreview2024}.

Beyond reducing the marginal cost of individual valuations, surrogates also offer a portfolio-level benefit: under natural consistency assumptions on model parameters and risk factors, the pricing operator is linear in positions, so portfolio values can be obtained by applying the surrogate directly and exploiting linearity, rather than building and evaluating separate surrogates instrument-by-instrument and then aggregating. This is particularly relevant in large-scale risk and scenario analysis, where portfolios may contain thousands of contracts and repeated revaluation is the dominant computational bottleneck.

One such surrogate is Gaussian Process Regression (GPR)~\cite{lehdili2019market,lehdili2025performance}. As a non-parametric method, it offers considerable flexibility while naturally quantifying predictive uncertainty. However, deploying GPR at scale is hindered by the cubic computational cost of kernel matrix factorization and quadratic memory growth with the number of training points, as well as by difficulties in capturing high-dimensional interactions without large datasets~\cite{rasmussen2006,ambikasaran2014,saatci2011}. Structured variants, such as sparse or inducing-point methods and separable kernels on grids, mitigate some of these issues but still struggle when the training set must grow to resolve complex, multi-asset American exercise boundaries~\cite{titsias2009,hensman2013,wilson2015,flaxman2015}. Beyond GPR, alternative approaches have included deep neural networks, which can learn expressive price surfaces but often require extensive hyperparameter tuning and large training datasets to generalize reliably~\cite{becker2019,ferguson2018}.

Tensor network approaches~\cite{Or_s_2014, oseledets2011} provide an efficient framework for representing and processing high-dimensional data, addressing key scalability limitations of GPR. They allow high-dimensional datasets to be efficiently represented and processed. In this paper, we investigate the use of tensor train (TT) approaches to price options. First, we build a TT representation of the training set using the TT-cross approximation (TCA)~\cite{oseledets2010ttcross} and then perform inference on the TT. To do so, we derive TT representations of a Laplacian kernel and its analytic (noise-free) inverse, enabling GP-style posteriors without iterative linear solves~\cite{kirstein2022}. We further show that the $L\to \infty$ limit recovers a practical TT-native multilinear interpolation scheme for off-grid evaluation. Empirically, on five-asset basket options, we observe substantially lower error at shorter training times than standard GPR for European options, enabled by scaling to much larger effective training sets. For American options, we match GPR-level accuracy with inference times in the millisecond range, while data generation (e.g., LSMC with 10{,}000 paths and 30 timesteps) dominates the wall-clock cost.

\paragraph{Paper organization.}
Section~2 details TT-cross for learning surrogates from black-box option pricers. Section~3 develops the STN-GPR formulation, including the Laplacian kernel and its analytic inverse in TT as well as TT-native interpolation for off-grid queries. Section~4 reports results for five-asset baskets: European puts on the geometric average (Fig.~1) and American puts on the arithmetic average (Fig.~2), covering error, training time, and dataset scaling. Section~5 concludes.

\section{TT-Cross approximation}
\label{sec:tt-cross}
Our first hurdle with the \gls{TT} approach is to represent the training vector, $\mathbf{y}$, as a \gls{TT}. We found that decomposing $\mathbf{y}$ via consecutive singular value decompositions yielded an efficient representation, but, for large training sets, even storing the $\mathbf{y}$ temporarily as a full vector becomes intractable. Fortunately, for both the use cases considered here, we have access to black-boxes which can generate option prices for a given set of parameters. This allows us to employ the \gls{TCA}. Here we outline the derivation, loosely following Ref.~\cite{oseledets2010ttcross}. 

First, the simpler matrix cross approximation applied to a matrix $A$ consists of
\begin{equation}\label{eq:matrixcross}
    A \approx A[:,\mathcal{I}](A[\mathcal{I},\mathcal{J}])^{-1}A[\mathcal{J}, :]=C\widehat{A}^{-1}R,
\end{equation}
where $\mathcal{I}$ and $\mathcal{J}$ correspond to some subset of column and row indices respectively. Here we have introduced $C$, $\widehat{A}$ and $R$ as the column matrix, intersection sub-matrix, and row matrix respectively. If the number of indices in $\mathcal{I}$ and $\mathcal{J}$ both equal the rank of $A$, then this becomes an equality. When this is not the case, the quality of the approximation comes down to how ``good'' the sub-matrix $A[\mathcal{I},\mathcal{J}]$ is.
It has been shown~\cite{goreinov2010how} that the ``best'' sub-matrix is the one with the maximum volume, i.e.,
\begin{equation}
    \widehat{A} = \max_{\mathcal{I},\mathcal{J}} \left| \det A[\mathcal{I},\mathcal{J}]\right|.
\end{equation}
This is an NP-hard problem and so we resort to an approximation. In this case, we can make use of the \texttt{maxvol} algorithm which iteratively fixes one of $\mathcal{I}$ or $\mathcal{J}$ and maximizes the volume over the other. 

Generalizing this to \gls{TT}s can be done in the following way. To begin, consider a $d$-dimensional tensor, $T_{i_1,i_2,\dots,i_d}$ with $i_k=\{1,n_k\}$, whose entries correspond to a function, $f$, evaluated over a discrete grid:
\begin{equation}
    T_{i_1,i_2,\dots,i_d}=f(i_1,i_2,\dots,i_d).
\end{equation}
This is not necessarily a function whose domain is of dimension $d$, as one or more dimensions could be encoded over multiple tensor indices.

The initial steps to introduce the TT-cross algorithm are purely analytical, in particular we will never actually store the full tensor. With this in mind, decomposing the tensor, $T$, into a \gls{TT} first involves reshaping it into a matrix of the form $T_{i_1,i_2 i_3\dots i_d}$, where the concatenation of indices corresponds to combining them into a single multi-index of dimension $\prod_{k=2}^d n_k$. Now, we wish to apply the decomposition of Eq.~\eqref{eq:matrixcross} to the reshaped matrix to arrive at
\begin{equation}
    T_{i_1,i_2 i_3\dots i_d} = C_{i_1,\alpha_1} \widehat{T}_{\alpha_1,\alpha_1}^{-1}R_{\alpha_1,i_2 i_3 \dots i_d},
\end{equation}
where we abuse notation to re-use $C$ and $R$ and we will denote the dimension of all $\alpha_i$ to $r_i$. The sub-matrix, $\widehat{T}$, is populated by the full tensor sampled at a left index set $\mathcal{I}_1=\{i_1^{(\alpha_1)}\}$ and a right index set $\mathcal{J}_1=\{j_2^{(\alpha_1)},j_3^{(\alpha_1)},\dots,j_d^{(\alpha_1)}\}$, i.e.,
\begin{equation}
\widehat{T}_{\alpha_1,\alpha_1}=T_{i_1^{(\alpha_1)},j_2^{(\alpha_1)}j_3^{(\alpha_1)}\dots j_d^{(\alpha_1)}}.
\end{equation}
Again, as with the matrix cross approximation, will apply the \texttt{maxvol} algorithm to optimize one of these index sets while leaving the other fixed. It is an easy choice: we begin by fixing $\mathcal{J}_1$ as $\mathcal{I}_1$ corresponds to a far smaller space over which to optimize. Initially, the right set is fixed to random values, and we can optimize over sub-matrices of size $r_1\times r_1$ within the matrix $C_{i_1,\alpha_1}$. Populating this matrix is the first numerical step of this algorithm, and each element is defined in the following way
\begin{equation}
    C_{i_1,\alpha_1} = f(i_1,j_2^{(\alpha_1)},j_3^{(\alpha_1)},\dots ,j_d^{(\alpha_1)}).
\end{equation}
Once done, we can contract $C\widehat{T}^{-1}$ into a single tensor, which shall be the first core of the \gls{TT}.

To split off the next core from the remainder, $R$, we first reshape this into a matrix of the form $R_{\alpha_1i_2,i_3i_4\dots i_d}$. Applying the cross decomposition now yields
\begin{equation}
    R_{\alpha_1 i_2,i_3i_4\dots i_d}=C_{\alpha_1 i_2,\alpha_2}\widehat{T}^{-1}_{\alpha_2,\alpha_2} R_{\alpha_2, i_3i_4\dots i_d}.
\end{equation}
We can then again find the maximum volume sub-matrix in $C_{\alpha_1 i_2, \alpha_2}$ whose entries are given by
\begin{equation}
    C_{\alpha_1 i_2, \alpha_2} = f(i_1^{(\alpha_1)},i_2,j_2^{(\alpha_2)},j_3^{(\alpha_2)},\dots ,j_d^{(\alpha_2)}),
\end{equation}
where the values of $i_1$ are those found in the optimization of the previous step. Each of the $r_2$ rows of the $\alpha_1 i_2$ multi-index identified by the optimization has an associated $i_1$ and $i_2$ index, these index pairs form the new left index set: $\mathcal{I}_2=\{i_1^{(\alpha_2)}, i_2^{(\alpha_2)}\}$.

At the start of step $n$ in the process we have the index sets
\begin{equation}
    \mathcal{I}_n = \{i_1^{(\alpha_{n-1})}, i_2^{(\alpha_{n-1})},\dots,i_{n-1}^{(\alpha_{n-1})}\}, \; \mathcal{J}_n = \{j_n^{(\alpha_n)},j_{n+1}^{(\alpha_n)},\dots,j_d^{(\alpha_n)}\}.
\end{equation}
After reaching and optimizing the final index, the process is repeated in the opposite direction. This proceeds identically to the previous sweep except the left and right index sets swap roles, and it is the right index set that is gradually grown with optimal indices added. The other main difference is that we now have a better initial guess for the indices in the left index set; we can simply use those found in the previous sweep.

\section{TT inference}
\label{sec:stn-gpr}

The \gls{TT} representation of a function is inherently restricted to a discrete domain. This restriction is not ideal in a pricing context unless the grid can be made sufficiently fine; when this is not the case, some form of interpolation is required. Interpolation may be performed using a trained model such as \gls{GPR}, or directly on the underlying \gls{TT} representation. We begin by outlining the main bottleneck of \gls{GPR}, before showing that the optimal choice of \gls{GPR} hyperparameters in our particular setting leads to a simple multilinear interpolation on the \gls{TT} data points.
\subsection{STN-GPR}
Bayesian inference is usually hindered by an intractable marginal likelihood; this is not the case with Gaussian Process Regression (GPR). The posterior distribution can be expressed analytically, but computing it requires solving linear systems involving the dense kernel matrix. For a dataset of size $n$, this entails $\mathcal{O}(n^3)$ time and $\mathcal{O}(n^2)$ memory, which quickly becomes prohibitive as $n$ grows. It is this computational bottleneck, rather than analytical intractability, that we seek to alleviate using tensor network methods.

While it is possible to address this issue using numerical approximation schemes, by decomposing a general kernel matrix into a \gls{TT} format and then applying an iterative algorithm~\cite{oseledets2012solution,dolgov2014alternating} to compute an approximate inverse, in this project we take a different approach. By choosing a specific kernel with an analytic inverse, we are able to directly construct a low-rank \gls{TT} representation.

The Laplacian kernel is defined as
\begin{equation}
    K(\mathbf{x},\mathbf{x}') = \exp(-|\mathbf{x}-\mathbf{x}'|_1/L)
\end{equation}
where $\{\mathbf{x},\mathbf{x}'\}$ are feature arrays, $|\cdot|_1$ is the Manhattan distance, and $L$ is the sole hyperparameter of the kernel and corresponds to the length-scale. A key assumption we make is that the training data for $N$ features lies on an $N$-dimensional hypercube lattice with uniform spacing and $2^n$ points along each dimension for some integer $n$, the lattice constant and value of $n$ for each feature can differ.

First, consider a single feature with lattice spacing $\Delta$ and $2^n$ training points, the kernel matrix in this case can be expressed as
\begin{equation}
    \mathbf{K}_n = \begin{pmatrix}
        1 & a & a^2 & \cdots & a^{2^n-1} \\
        a & 1 & a & \cdots & a^{2^n-2}\\ 
        a^2 & a & 1 & \cdots & a^{2^n-3}\\
        \vdots & \vdots & \vdots & \ddots & \vdots\\
        a^{2^n-1} & a^{2^n-2} & a^{2^n-3} & \cdots & 1
    \end{pmatrix},
\end{equation}
where $a=\exp(-\Delta/L)$. To find the \gls{TT} representation of this matrix we first write
\begin{equation}\label{eq:bowtieintro}
    \mathbf{K}_n = \begin{pmatrix}
        \mathbb{I}_2 & \sigma^+ & \sigma^-
    \end{pmatrix} \bowtie
    \begin{pmatrix}
        \mathbf{K}_{n-1} \\
        \mathbf{M}_{n-1} \\
        \mathbf{M}_{n-1}^T
    \end{pmatrix},
\end{equation}
where have introduced the following matrices for convenience:
\begin{equation}
    \mathbb{I}_2 = \begin{pmatrix}
        1 & 0 \\
        0 & 1
    \end{pmatrix}, \hspace{0.2cm}
    \sigma^+ = \begin{pmatrix}
        0 & 1 \\
        0 & 0
    \end{pmatrix} = (\sigma^-)^T, \hspace{0.2cm}
    \mathbf{M}_n = \begin{pmatrix}
        a^{2^n} & a^{2^n+1} & \cdots & a^{2^{n+1}-1} \\
        a^{2^n-1} & a^{2^n} & \cdots & a^{2^{n+1}-2} \\
        \vdots & \vdots & \ddots & \vdots \\ 
        a & a^2 & \cdots & a^{2^n}.
    \end{pmatrix}
\end{equation}
Here the bowtie product ($\bowtie$) corresponds to a matrix product along the outer index and a tensor product of the inner indices, i.e., we could also write Eq.~\eqref{eq:bowtieintro} as
\begin{equation}
    \mathbf{K}_n = \mathbb{I}_2 \otimes \mathbf{K}_{n-1} + \sigma^+ \otimes \mathbf{M}_{n-1}+ \sigma^- \otimes \mathbf{M}_{n-1}^T.
\end{equation}

We can in turn decompose $\mathbf{M}_n$ as follows
\begin{equation}
    \mathbf{M}_n = 
        (a^{2^{n-1}}\mathbb{I}_2 + a^{2^n} \sigma^+ + \sigma^-)\otimes \mathbf{M}_{n-1}.
\end{equation}
Iterating the above decompositions allow us to express the kernel as
\begin{equation}
    \mathbf{K}_n = \begin{pmatrix}
        \mathbb{I}_2 & \sigma^+ & \sigma^-
    \end{pmatrix} \bowtie \mathbf{C}_{n-2} \bowtie \mathbf{C}_{n-3}\bowtie \cdots \bowtie \mathbf{C}_1 \bowtie \begin{pmatrix}
        \mathbf{K}_1 \\ \mathbf{M}_1 \\ \mathbf{M}_1^T
    \end{pmatrix},
\end{equation}
where we have define the central cores $\mathbf{C}_i$ given by
\begin{equation}
    \mathbf{C}_i = \begin{pmatrix}
        \mathbb{I}_2 &\sigma^+ & \sigma^- \\
        0 & a^{2^i}\mathbb{I}_2 + a^{2^{i+1}} \sigma^+ + \sigma^- & 0 \\
        0 & 0 & a^{2^i}\mathbb{I}_2 + a^{2^{i+1}} \sigma^- + \sigma^+
    \end{pmatrix}.
\end{equation}
Hence $\mathbf{K}_n$ can be exactly represented by a rank-3 tensor train and, through abuse of notation, the $0$ entries are $2\times 2$ matrices of zeros.

Without noise this kernel can be inverted analytically, considering a single feature with the same training grid as before we have
\begin{equation}
    \mathbf{K}_n^{-1} = \frac{1}{1-a^2}\begin{pmatrix} 1 &-a & & & &  \\
    -a & 1+a^2 & -a & &  & \\
     & -a & 1+a^2 & \ddots & &    \\
     &  & \ddots & \ddots & -a &   \\
     &  &  &  -a & 1+a^2 & -a  \\
     & & & &  -a &1
        
    \end{pmatrix}.
\end{equation}
Through a similar analysis as with the kernel itself, the inverse can be represented as a rank-5 TT.

Extending all of the above to multiple features is trivial as the multi-dimensional Laplacian kernel permits a tensor product structure provided the training samples are suitably ordered. As an example, consider two features with lattice spacings $\Delta_1$ and $\Delta_2$ and just 2 points along each lattice. Then, defining $a=\exp(-\Delta_1/L)$ and $b=\exp(-\Delta_2/L)$, we can write the kernel matrix as
\begin{equation}
    \mathbf{K}_1 = \begin{pmatrix}
        1 & b & a & ab \\
        b & 1 & ab & a \\ 
        a & ab & 1 & b \\
        ab & a & b & 1
    \end{pmatrix}= \begin{pmatrix}
        1 & a \\
        a & 1
    \end{pmatrix} \otimes \begin{pmatrix}
        1 & b \\
        b & 1
    \end{pmatrix} = \mathbf{K}_1^{(a)}\otimes\mathbf{K}_1^{(b)}.
\end{equation}
This factorization is possible thanks to the Manhattan distance and means we can build a \gls{TT} representation of the kernel and its inverse for each feature separately and connect them with tensor products.

To infer the price at some points, $\mathbf{X}_*$, not in the training grid we still require a \gls{TT} representation for $K(\mathbf{X}_*, \mathbf{X})$. For simplicity, we will only consider inferring one point, $x_*$, i.e., we wish to compute $K(x_*,\mathbf{X})$. This corresponds to a vector of length $2^n$. We can again construct a \gls{TT} representation of this vector one core at a time and the resulting \gls{TT} vector is rank-3.

We haven't accounted for noise in any of the above, doing so would require \gls{TT} representations of $\mathbf{K}+\sigma^2\mathbb{I}$ and its inverse. The former is simple, as the identity is rank 1 and the addition of \gls{TT}s is an established routine, however, we no longer have an analytical expression for the inverse. Therefore, if we wished to include noise, we would need to use a numerical approximation to compute the inverse. In practice, we found this to be too computationally expensive to achieve a reasonable error.

Even in the absence of observation noise, the model retains a nontrivial hyperparameter in the form of the kernel length-scale $L$. Empirically, in our experiments we optimized $L$ by minimizing the negative log marginal likelihood and found that the optimizer consistently drifted toward increasingly large values of $L$. Within this model class, larger $L$ corresponds to longer-range prior correlations and therefore favors slowly varying interpolants among all functions that exactly match the data.

This observation motivates examining the large-$L$ regime (with $\sigma=0$), in which posterior inference simplifies substantially. In one dimension, the Gaussian process prior with the Laplacian kernel coincides with the covariance structure of a stationary Ornstein-Uhlenbeck (OU) process~\cite{sarkka2019applied}. Given this, conditioning on function values at two locations reduces inference on the interval between them to computing the conditional mean of an OU bridge. For $x_0<x^*<x_1$, this conditional mean admits the closed-form expression~\cite{barczy2013representations}
\begin{equation}\label{eq:ouweights}
y^* = y_0 \frac{\sinh \!\left(\frac{x_{1}-x^*}{L}\right)}{\sinh\!\left(\frac{x_1-x_0}{L}\right)} \;+\; y_1 \frac{\sinh \!\left(\frac{x^*-x_0}{L}\right)}{\sinh\!\left(\frac{x_1-x_0}{L}\right)}.
\end{equation}
Taking $L\to\infty$ and using $\sinh(z/L)\sim z/L$ yields
\begin{equation}
y^* \to y_0\,\frac{x_1-x^*}{x_1-x_0} + y_1\,\frac{x^*-x_0}{x_1-x_0},
\end{equation}
i.e., standard linear interpolation. While the prior covariance becomes degenerate in the strict $L\to\infty$ limit (the constant-kernel limit), the limiting posterior mean obtained from finite $L$ recovers linear interpolation.

Given the tensor product structure of the Laplacian kernel, posterior computations (mean, variance, and linear solves) inherit the same tensor-product algebra. In particular, when conditioning on values at the corners of an axis-aligned hyperrectangle, the interpolation weights factorize into products of the corresponding one-dimensional bridge weights given in Eq.~\eqref{eq:ouweights}; taking the large-$L$ limit then yields the standard multilinear interpolation rule. We will now cover how this permits an efficient representation in the \gls{TT} representation.
\subsection{STN-interpolation}
Given two data points, $(x_0,y_0)$ and $(x_1,y_1)$, we can approximate the y-coordinate of a point $(x^*,y^*)$ where $x_0\leq x^* \leq x_1$ using linear interpolation via
\begin{equation}\label{eq:linearint}
    y^* = y_0\left(\frac{x_1-x^*}{x_1-x_0}\right) + y_1\left(\frac{x^*-x_0}{x_1-x_0}\right).
\end{equation}
Now, assume we have a set of $n=2^m$ data points $\{(x_i,y_i)\}$ with $x_0 \leq x_1 \leq \dots \leq x_{n}$ and we store the $y$-values in a vector, $\mathbf{y}$. Interpolating a point $y^*$ located at $x^*$, where $x_0\leq x^* \leq x_{n}$, can be expressed as
\begin{equation}\label{eq:linintdot}
    y^* = \mathbf{v}^T_{x^*}\mathbf{y},
\end{equation}
where $\mathbf{v}_{x^*}$ contains interpolation coefficients such as those in Eq.~\eqref{eq:linearint}.

If $\mathbf{y}$ is especially large it may be desirable to represent this as a \gls{TT} and carry out Eq.~\eqref{eq:linintdot} purely in terms of TTs. This requires finding a \gls{TT} representation for $\mathbf{v}_{x^*}$ as well, fortunately because this only has two non-zero elements it can't have a rank larger than 2. The position of these two elements within the vector is key to determining the actual structure of the representation. For instance, if the first two elements of $\mathbf{v}_{x^*}$ are non-zero then it can be expressed as a rank-1 \gls{TT}:
\begin{equation}
    \mathbf{v}_{x^*} = \begin{pmatrix}
        c_0 \\
        c_1 \\
        0 \\
        0 \\
        \vdots
    \end{pmatrix} = \mathbf{e}_1 \otimes \mathbf{e}_1 \otimes \dots \otimes \mathbf{e}_1 \otimes \begin{pmatrix}
        c_0 \\
        c_1
    \end{pmatrix},
\end{equation}
where we have introduced the local basis $\mathbf{e}_1=\begin{pmatrix} 1 \\ 0 \end{pmatrix}$ (along with $\mathbf{e}_0=\begin{pmatrix} 0 \\ 1 \end{pmatrix}$). Compared to this, if the coefficients are located exactly in the centre then the ranks throughout are 2:
\begin{equation}
    \mathbf{v}_{x^*} = \begin{pmatrix}
        0 \\
        0 \\
        \vdots \\
        c_0 \\
        c_1 \\
        \vdots \\ 
        0 \\ 
        0
    \end{pmatrix} = \begin{pmatrix}
        \mathbf{e}_1 & \mathbf{e}_0
    \end{pmatrix} \bowtie \begin{pmatrix}
        \mathbf{e}_0 & 0 \\
        0 & \mathbf{e}_1 
    \end{pmatrix} \bowtie \begin{pmatrix}
        \mathbf{e}_0 & 0 \\
        0 & \mathbf{e}_1 
    \end{pmatrix} \bowtie \cdots \bowtie
    \begin{pmatrix}
        \mathbf{e}_0 & 0 \\
        0 & \mathbf{e}_1 
    \end{pmatrix} \bowtie 
    \begin{pmatrix}
        c_0 \mathbf{e}_0 \\
        c_1 \mathbf{e}_1
    \end{pmatrix}.
\end{equation}
Here the subscripts of the basis vectors are a binary string encoding the position of each coefficient. The first example is rank 1 because the two strings are identical down to the last bit, whereas in the second example every bit is different. For a generic pair of neighbouring strings the \gls{TT} will be some mix of rank 1 and rank 2 cores. To build the \gls{TT} for a generic $\mathbf{v}_{x^*}$, first encode the index of each coefficient as a binary string, store these and record the output of applying an \texttt{AND} operation. As these are neighbouring strings, the resulting string of the \texttt{AND} operation will be some number (perhaps none) of 1's followed by a region of purely 0's. Where this crossover occurs determines where the \gls{TT} rank will increase from 1 to 2.

For multi-dimensional interpolation this is most simply extended by considering a multilinear/tensor product interpolation scheme. Here the coefficient \gls{TT} is a tensor product of a coefficient \gls{TT} for each dimension. Note that we haven't specified the explicit form of $c_0$ and $c_1$, so while this applies for linear interpolation, it also could be used, for example, for the more general hyperbolic-sine interpolation of Eq.~\eqref{eq:ouweights}

\subsection{Direct Portfolio Valuation via Linearity}

In practical risk management applications, valuation is typically performed at the portfolio level rather than instrument by instrument. Consider a portfolio composed
of instruments with prices $P_i(x)$ depending on a vector of market risk factors
$x \in \mathbb{R}^d$, and portfolio weights $w_i \in \mathbb{R}$. The portfolio value
is given by
\begin{equation}
V(x) = \sum_{i=1}^{M} w_i P_i(x).
\end{equation}
Pricing is therefore linear in the instrument values.

A key structural advantage of the tensor-train (TT) representation is that it is
closed under linear combinations (up to controlled rank growth). If each
$P_i(x)$ admits a TT representation,
\begin{equation}
P_i(x) \approx \mathcal{T}_i(x),
\end{equation}
then the portfolio value can be represented as
\begin{equation}
V(x) \approx \sum_{i=1}^{M} w_i \mathcal{T}_i(x),
\end{equation}
which itself can be compressed back into TT form. Consequently, inference may be performed directly at the portfolio
level without requiring option-by-option aggregation at evaluation time.

Alternatively, when instruments share a consistent set of model parameters and
risk-factor grids, one may construct a portfolio-level surrogate directly by applying
TT-cross to the aggregate pricing functional
\begin{equation}
V(x) = \sum_{i=1}^{M} w_i P_i(x),
\end{equation}
using calls to a trusted portfolio pricer. 
This setting avoids duplicating surrogate models across trades and ensures that evaluation
cost remains essentially independent of the number of instruments once the TT
representation has been constructed.

These properties are particularly relevant in large-scale market risk simulations,
where the dominant computational burden arises from repeated portfolio
revaluation under extensive scenario sets.

\section{Results}
\label{sec:res}
In this section, we assess the performance of classical \gls{GPR} and \gls{STN} models for pricing put options written on the geometric average of a basket of five assets. Both models are trained on a dataset comprising eight input features, including spot prices, strike prices, interest rates, and time to maturity. These features are discretized on a high-dimensional grid to adequately capture the complex structure of the option pricing surface and the underlying market dynamics.

The models are trained over the following feature ranges:
\begin{itemize}
    \item Spot prices for each asset in the range \$5 to \$150,
    \item Option strike prices ranging from \$1 to \$200,
    \item Interest rates between 0.5\% and 8\%,
    \item Times to maturity (TTM) spanning from 1 day to 3 years.
\end{itemize}
Altogether, this yields an eight-dimensional input space. For the \gls{TT}-based approach, the discretization of each feature must be specified explicitly. We found the following grid configuration to provide a favorable trade-off between accuracy and computational efficiency:
\begin{equation}\label{eq:grid}
    \underbrace{32\times32\times32\times32\times32}_{\text{Spot prices}} \times \underbrace{64}_{\substack{\text{Strike}\\ \text{price}}} \times \underbrace{8}_{\substack{\text{Interest}\\\text{rate}}} \times \underbrace{8}_{\text{TTM}},
\end{equation}
which corresponds to a tensor train representation with 37 cores, each of local dimension 2.

The \gls{TCA} is initialized using a state obtained via the TT-ANOVA algorithm~\cite{chertkov2023black} up to second order with rank 2. For brevity, the implementation details of this initialization procedure are omitted. Importantly, the reported \gls{TT} results account for both the computational cost and the sample complexity associated with the TT-ANOVA step.

As a high-fidelity benchmark, we employ \gls{LSMC} simulations. Model accuracy is evaluated using the \gls{MAE}, computed on independent test sets of 1{,}000 samples uniformly drawn from the prescribed parameter space. For the TT results we made use of the \texttt{torchTT}~\cite{torchtt2025} and \texttt{teneva}~\cite{teneva2025} libraries, for the GPR results we made use of the \texttt{scikit-learn}~\cite{scikit-learn} and for the LSMC simulations we used \texttt{QuantLib}~\cite{quantlib2025}.

While we report errors at the level of individual contracts, the intended application is portfolio revaluation, where such surrogates would be embedded in large-scale VaR or ES calculations.

\subsection{European basket option}
\begin{figure}[h!]
    \centering
    \includegraphics[width=0.48\linewidth]{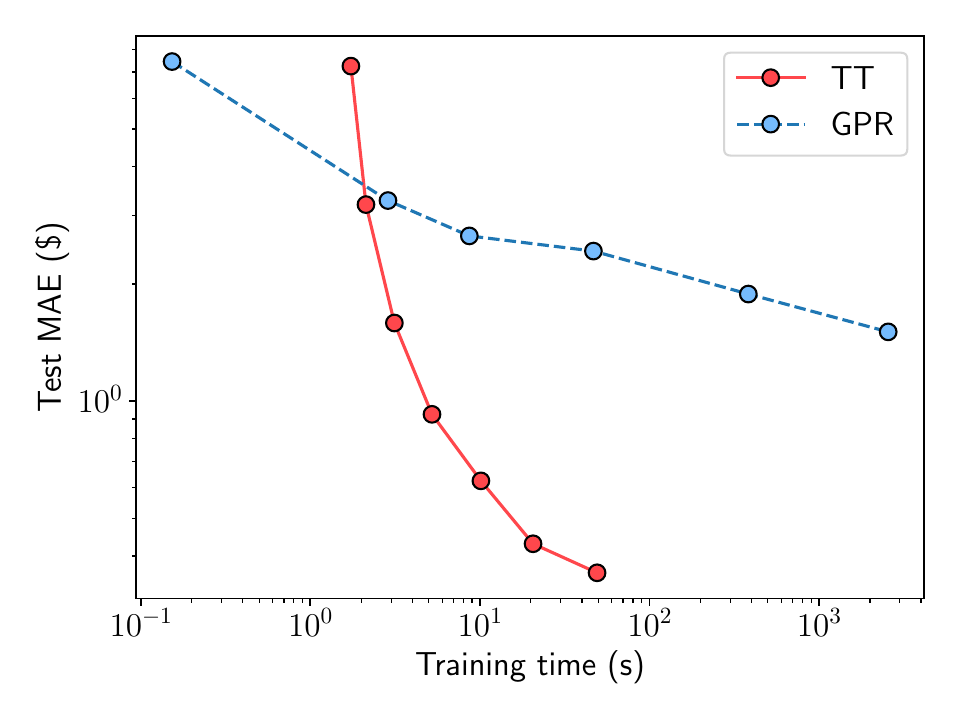}
    \includegraphics[width=0.48\linewidth]{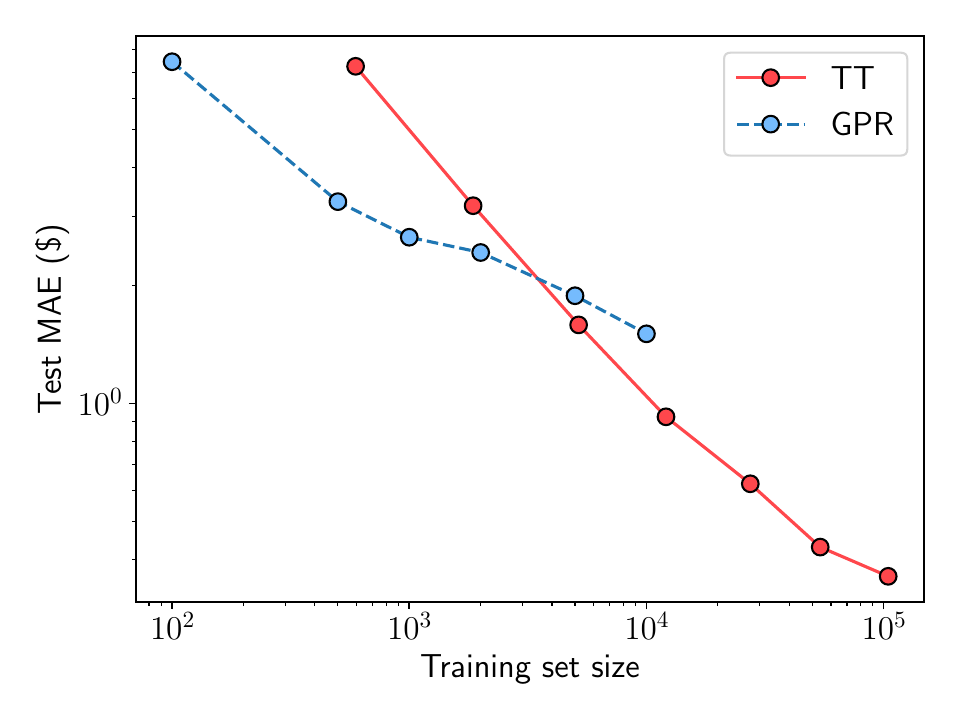}
    \caption{Error in pricing a European put geometric basket option. Comparison in test error of TT-interp and \gls{GPR} approaches as a function of the training time for varying training set sizes.}
    \label{fig:eu_comp}
\end{figure}

In this section we compare the performance of the \gls{GPR} and \gls{STN} approaches in modelling the price of a European put option on the geometric average of a basket of five assets. We plot the results of this comparison in Figure~\ref{fig:eu_comp} where we see in the left-hand plot that the \gls{STN} approach is able to achieve a much lower error than the \gls{GPR} approach at significantly shorter training times. In the right-hand plot we see that this is possible due to the \gls{STN} approach allowing access to much larger training set sizes. To be clear, in the \gls{STN} approach the training set corresponds to those function evaluation used in building the \gls{TCA} of the function evaluated over the grid defined in Eq.~\eqref{eq:grid}. So, in a sense, two interpolations are being performed: the \gls{TCA} interpolates between the function evaluations and the subsequent TT-interpolation interpolates between the grid points. To highlight the power of this approach, note that the grid as a whole contains 137.5bn points, whereas the most function evaluations we use in the \gls{TCA} of the function over these grid points is 500k. For completeness, we attempted to push the \gls{TT} approach to larger training sets but began to see instabilities in the \gls{TCA} meaning we lost the monotonic increase in accuracy. We have omitted these results but believe that with suitable hyperparameter tuning of the \gls{TCA} this could be resolved, we leave this for a future work.

\subsection{American basket option}
The price of a European option has a closed-form solution. This is not the case for American options; the potential for early exercise makes an analytical solution intractable and necessitates a numerical approach. Of these approaches, the two most prominent are the binomial tree (BT) and \gls{LSMC} algorithms. Both methods can accurately approximate American option prices by explicitly accounting for the early exercise feature, but incur significant computational costs. These costs are exacerbated by the presence of multiple assets and/or the necessity to compute many prices for, e.g., a risk calculation~\cite{lehdili2019market,lehdili2025performance}.

Here we seek to avoid this cost by training a model on the output of \gls{LSMC} for an American put option on the arithmetic average of a basket of five assets. We compare two models, classical \gls{GPR} and a quantum-inspired \gls{STN} approach, and show that pricing via evaluation of the trained models is significantly cheaper than evaluating the \gls{LSMC} algorithm. For the LSMC algorithm we used 10000 samples and 30 timesteps.

\begin{figure}[h!]
    \centering
    \includegraphics[width=0.48\linewidth]{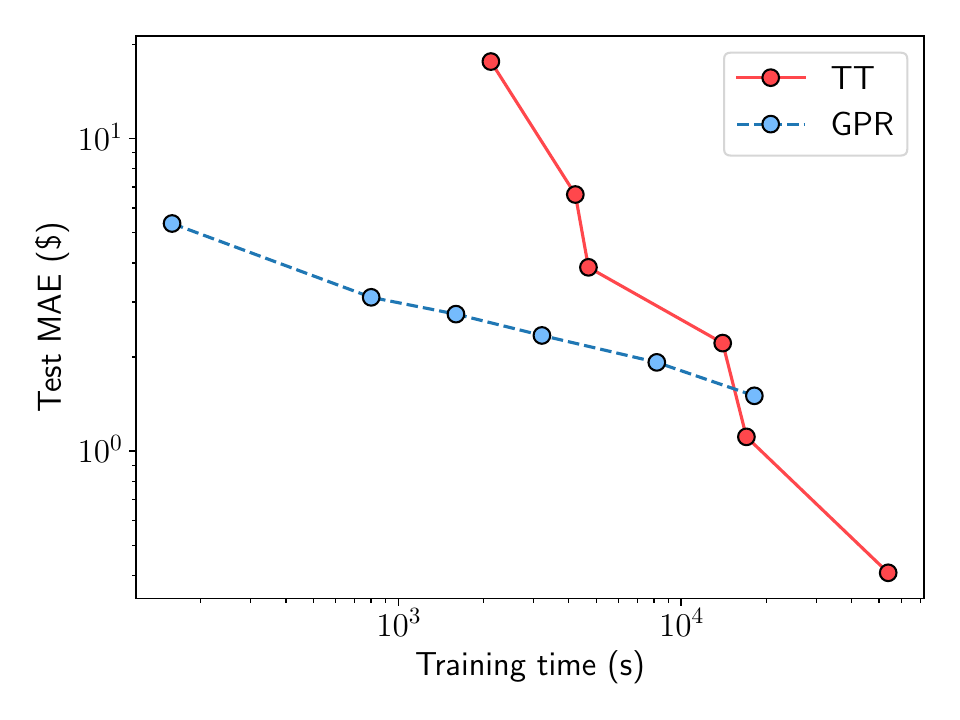}
    \includegraphics[width=0.48\linewidth]{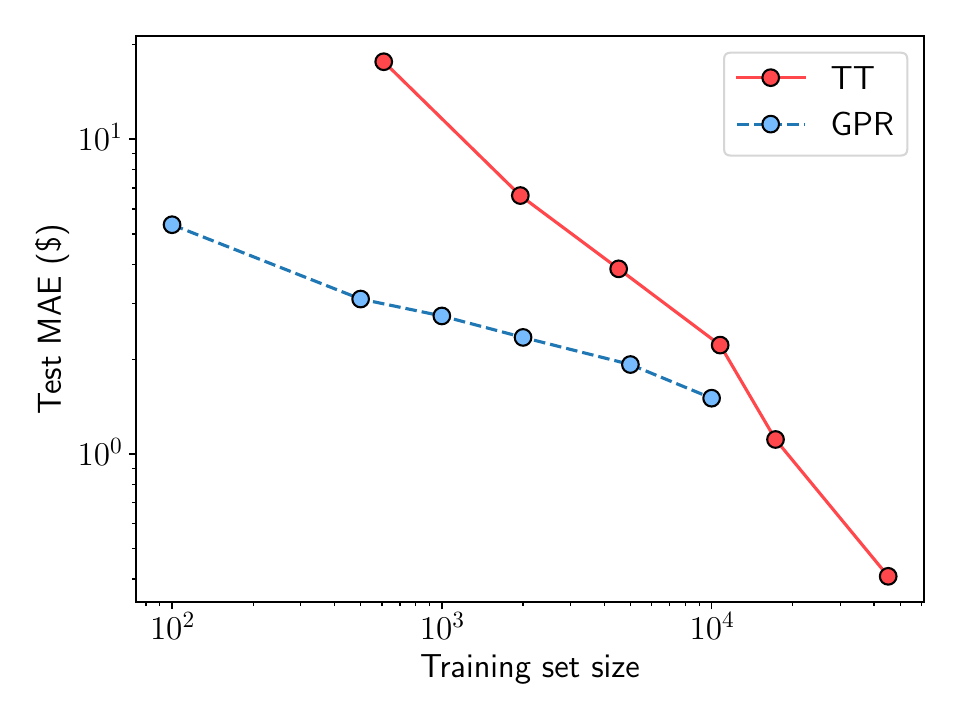}
    \caption{Error in pricing a American put arithmetic basket option. Comparison in test error of TT-interp and \gls{GPR} approaches as a function of the training time for varying training set sizes. Here the training time includes both the time taken to generate the dataset and, for \gls{GPR}, the time taken to tune hyperparameters.}
    \label{fig:american_comp}
\end{figure}

In Figure \ref{fig:american_comp} we compare the test error of the \gls{STN} and \gls{GPR} approaches as a function of training time for varying training set sizes. As in the European option case, the \gls{STN} approach can be trained on larger datasets and thereby reach lower errors than exact \gls{GPR}. In this American option experiment, however, the overall wall-clock time is dominated by data generation, since each \gls{LSMC} run takes $\sim 1$s, and therefore the measured training times for \gls{STN} and \gls{GPR} are similar over the range considered. We nevertheless cap exact \gls{GPR} at $N=10^4$ training samples because its computational and memory requirements scale unfavorably with dataset size (training $\mathcal{O}(N^3)$ and inference $\mathcal{O}(N^2)$), so pushing to substantially larger $N$ would quickly become impractical for the intended setting of repeated pricing. In contrast, once trained, \gls{STN} evaluation is roughly independent of $N$ and remains cheap per query. For the training set sizes considered here, inference takes around $10^{-3}$s per sample for \gls{STN} and $10^{-4}$s for \gls{GPR}, but \gls{STN} can be scaled to larger training sets in this setting to further reduce the approximation error.

\section{Conclusions}
\label{sec:conclusion}
We introduced a tensor network surrogate model for option pricing that combines \gls{TT}-cross approximation, analytic \gls{TT} kernels, and \gls{TT}-native interpolation. Focusing on American-style features, the proposed approach amortizes the cost of an expensive black-box pricer into a compact representation that can be evaluated extremely rapidly for downstream tasks such as scenario analysis and repeated revaluation. By exploiting separability and low rank structure, the method scales to high dimensional input spaces and large training grids that are prohibitive for classical \gls{GPR}, while preserving a kernel-based interpretation and associated uncertainty quantification.

Beyond pointwise accuracy and speed, an important advantage of the proposed approach lies in its suitability for repeated pricing contexts. In many risk management applications, the same instrument or portfolio must be revalued thousands to millions of times in order to compute tail risk measures such as Value-at-Risk (VaR), Expected Shortfall (ES), or to perform stress testing~\cite{lehdili2019market,lehdili2025performance}. In these settings, the computational burden is dominated by the cost of repeated pricing calls, making surrogate models particularly attractive. Recent work has demonstrated that Gaussian process based surrogates can significantly accelerate market risk calculations without compromising accuracy, especially in trading book applications.

A natural direction for future work is therefore to embed the proposed \gls{STN}-\gls{GPR} framework directly into a market risk pipeline. In particular, it would be highly relevant to conduct a systematic performance comparison between \gls{STN}-\gls{GPR}, classical \gls{GPR}, and direct pricing using standard numerical methods such as \gls{LSMC}, finite difference or finite element PDE solvers, and tree-based methods. Such a study could be carried out within a VaR or ES computation, assessing both the accuracy of the resulting risk measures and the overall computational efficiency. This would provide a concrete demonstration of the operational benefits of tensor network surrogates in a realistic risk management setting.

Several additional extensions are also of interest. These include the incorporation of observation noise into the \gls{STN} framework, adaptive sampling strategies that focus training points in regions of high sensitivity such as near early exercise boundaries, and the use of richer kernel families beyond the Laplacian kernel considered here. Finally, given recent advances in Gaussian process based acceleration of credit and counterparty risk calculations, it would be interesting to explore whether similar tensor network techniques could be applied to repeated valuation problems arising in XVA and related settings~\cite{lehdili2024leveraging}.

\section{Acknowledgement}
The authors acknowledge the financial support of the Île-de-France Region through the Pack Quantique programme, under Programme No. 23007011.

\bibliographystyle{unsrt}
\bibliography{references}
\end{document}